\documentclass[a4paper,11pt]{article}
\pdfoutput=1

\usepackage{amssymb,mathrsfs,amsmath}
\usepackage{a4wide}
\usepackage{color,xcolor}
\usepackage{slashed,soul}
\usepackage{graphicx}
\usepackage{amsfonts}
\usepackage{lscape}
\def\linkcolor{cyan!70!black}

\usepackage[
colorlinks=true
,urlcolor=\linkcolor
,anchorcolor=\linkcolor
,citecolor=\linkcolor
,filecolor=\linkcolor
,linkcolor=\linkcolor
,menucolor=\linkcolor
,linktocpage=true
,pdfproducer=medialab
,pdfa=true
]{hyperref}
\usepackage{amsthm}
\usepackage{booktabs}
\usepackage{array}
\usepackage{rotating}
\usepackage[numbers, sort&compress]{natbib}
\usepackage{multirow}
\usepackage{float}
\usepackage[utf8]{inputenc}
\usepackage[T1]{fontenc}
\usepackage{appendix}
\usepackage{epsfig}
\usepackage{dsfont}

\newcommand{\be}{\begin{equation}}
\newcommand{\ee}{\end{equation}}

\newcommand{\beq}{\begin{equation}} 
\newcommand{\eeq}{\end{equation}} 
\newcommand{\ba}{\begin{array}}  
\newcommand{\ea}{\end{array}} 
\newcommand{\bea}{\begin{eqnarray}}  
\newcommand{\eea}{\end{eqnarray} }  
\newcommand{\bal}{\begin{align}}
\newcommand{\eal}{\end{align}}   
\newcommand{\bi}{\begin{itemize}}  
\newcommand{\ei}{\end{itemize}}  
\newcommand{\ben}{\begin{enumerate}}  
\newcommand{\een}{\end{enumerate}}  
\newcommand{\bc}{\begin{center}}
\newcommand{\ec}{\end{center}} 
\newcommand{\bt}{\begin{table}}
\newcommand{\et}{\end{table}}  
\newcommand{\btb}{\begin{tabular}}
\newcommand{\etb}{\end{tabular}}

\renewcommand{\baselinestretch}{1.2}

\allowdisplaybreaks

\let\OLDthebibliography\thebibliography
\renewcommand\thebibliography[1]{
  \OLDthebibliography{#1}
  \setlength{\parskip}{0pt}
  \setlength{\itemsep}{0pt plus 0.3ex}
}

\begin{document}


\begin{titlepage}

\vspace*{-3.5truecm}
\begin{flushright}
IFT-UAM/CSIC-22-52 
 \end{flushright}
\vspace{0.2truecm}

\begin{center}
\renewcommand{\baselinestretch}{1.8}\normalsize
\boldmath
{\LARGE\textbf{
A method for approximating optimal statistical significances with machine-learned likelihoods
}}
\unboldmath
\end{center}

\vspace{0.4truecm}

\renewcommand*{\thefootnote}{\fnsymbol{footnote}}

\begin{center}

{\bf 
Ernesto Arganda,$\,^{a,b}$\footnote{\href{mailto:ernesto.arganda@csic.es}{ernesto.arganda@csic.es}}, 
Xabier Marcano,$\,^{a,c}$\footnote{\href{mailto:xabier.marcano@uam.es}{xabier.marcano@uam.es}}
Víctor Martín Lozano,$\,^{d,e}$\footnote{\href{mailto:victor.lozano@desy.de}{victor.lozano@desy.de}}
Anibal D. Medina,$\,^{b}$\footnote{\href{mailto:anibal.medina@fisica.unlp.edu.ar}{anibal.medina@fisica.unlp.edu.ar}}
Andres D. Perez,$\,^{b}$\footnote{\href{mailto:andres.perez@iflp.unlp.edu.ar}{andres.perez@iflp.unlp.edu.ar}}
Manuel Szewc$\,^{f,g}$\footnote{\href{mailto:manuel.szewc@ijs.si}{manuel.szewc@ijs.si}}
and Alejandro Szynkman$\,^{b}$\footnote{\href{mailto:szynkman@fisica.unlp.edu.ar}{szynkman@fisica.unlp.edu.ar}}}

\vspace{0.5truecm}

{\footnotesize

$^a${\sl Instituto de Física Teórica UAM-CSIC, \\ C/ Nicol\'as Cabrera 13-15, Campus de Cantoblanco, 28049, Madrid, Spain \vspace{0.15truecm}}

$^b${\sl  IFLP, CONICET - Dpto. de Física, Universidad Nacional de La Plata, \\ C.C. 67, 1900 La Plata, Argentina \vspace{0.15truecm}}

$^c${\sl Departamento de Física Teórica, Universidad Autónoma de Madrid ,\\ E-28049 Cantoblanco, Madrid, Spain\vspace{0.15truecm}}

$^d${\sl Departament de Física Teòrica and IFIC, Universitat de València-CSIC, \\ E-46100, Burjassot, Spain \vspace{0.15truecm}}

$^e${\sl Deutsches Elektronen-Synchrotron DESY, Notkestr. 85, 22607 Hamburg, Germany \vspace{0.15truecm}}

$^f${\sl Jo\v zef Stefan Institute, Jamova 39, 1000 Ljubljana, Slovenia \vspace{0.15truecm}}

$^g${\sl International Center for Advanced Studies (ICAS) and ICIFI, UNSAM, \\ Campus Miguelete, 25 de Mayo y Francia, CP1650, San Martín, Buenos Aires, Argentina \vspace{0.15truecm}}
}

\vspace*{2mm}

\end{center}

\renewcommand*{\thefootnote}{\arabic{footnote}}
\setcounter{footnote}{0}

\begin{abstract}
\noindent 
Machine-learning techniques have become fundamental in high-energy physics and, for new physics searches, it is crucial to know their performance in terms of experimental sensitivity, understood as the statistical significance of the signal-plus-background hypothesis over the background-only one. 
We present here a simple method that combines the power of current machine-learning techniques to face high-dimensional data with the likelihood-based inference tests used in traditional analyses, which allows us to estimate the sensitivity for both discovery and exclusion limits through a single parameter of interest, the signal strength.
Based on supervised learning techniques, it can perform well also with high-dimensional data, when traditional techniques cannot.
We apply the method to a toy model first, so we can explore its potential, and then to a LHC study of new physics particles in dijet final states.
Considering as the optimal statistical significance the one we would obtain if the true generative functions were known, we show that our method provides a better approximation than the usual naive counting experimental results.
\end{abstract}

\end{titlepage}

\tableofcontents

\section{Introduction}

Machine learning (ML) techniques have become basic tools for data analysis in recent years, and particle physics is no exception. In fact, ML algorithms are playing a fundamental role in collider physics (for seminal papers see, for instance,~\cite{Denby:1987rk,Lonnblad:1990bi,Baldi:2014kfa} and for recent reviews see~\cite{Larkoski:2017jix,Guest:2018yhq,Albertsson:2018maf,Radovic:2018dip,Carleo:2019ptp,Bourilkov:2019yoi,Karagiorgi:2021ngt,Feickert:2021ajf}) and are already used practically as a standard tool in the experimental LHC searches carried out by the ATLAS and CMS collaborations (see, for instance,~\cite{ATLAS:2017bqn,ATLAS:2018tgm,ATLAS:2019fxb,CMS:2019uxx,CMS:2020poo,ATLAS:2020efs,ATLAS:2020pvn,CMS:2020uim,CMS:2021kom}). These ML methods can also be applied at the ensemble level of data~\cite{Khosa:2019kxd,Mullin:2019mmh,Chang:2020rtc,Flesher:2020kuy,Lai:2020byl,Arganda:2021azw} and it has been demonstrated that, under the assumption of independent and identically distributed events, one can construct the optimal multi-event classifier from a single-event classifier and, moreover, that these multi-event classifiers give rise to optimal single-event classifiers~\cite{anthony_bartlett_1999,Goodfellow-et-al-2016}, see also Refs.~\cite{Metodiev:2017vrx,Nachman:2021yvi} for some examples applied in the HEP field.

Once we have a trained classifier, the fundamental question from the point of view of the search for beyond Standard Model (BSM) physics is how to quantify its performance in terms of experimental sensitivities. 
The most extended use of such a classifier considers a cut or working point (WP) of its output, which defines a region that ideally favors the signal over the background, and then it computes the sensitivities following standard techniques~\cite{Cowan:2010js} taking into account only those events passing this cut. 
Nevertheless, this is nothing but a refined procedure of defining a signal region where to perform a search, and one might still wonder whether it is possible to directly connect the ML classifiers with the standard statistical tests, using its output in full glory and without the need of defining a working point. Experimental collaborations such as ATLAS and CMS do have a method to incorporate the full output distribution to a larger extent, see {\it e.g.}~Ref.~\cite{ATLAS:2012byx}\footnote{We thank Pietro Vischia and Sergio Sánchez Cruz for pointing the existing literature to us.}. They treat the classifier output simply as a better variable to bin and perform a Binned Likelihood fit on. Although powerful, this treatment can be unsatisfactory as it washes over the probabilistic interpretation of the trained classifier. This is evident in the incorporation of systematic uncertainties which are propagated to the classifier output as it would be for any other high-level observable, without any re-training. This strategy means that the learned classifier is not necessarily a monotonous function of the Likelihood Ratio and perhaps the obtained significance is a sub-optimal approximation of the achievable significance. An analysis strategy that incorporates the full probabilistic structure of the classifier would perhaps be a more natural fit for incorporating systematic uncertainties and would be able to guarantee a better approximation to the full Likelihood ratio.

Some work in this direction has been recently done, see {\it e.g.}~\cite{Cranmer:2015bka, Elwood:2018qsr,DAgnolo:2018cun,Nachman:2019dol,DAgnolo:2019vbw,Chen:2020mev,Matchev:2020wwx,Cornell:2021gut,Aguilar-Saavedra:2021utu,dAgnolo:2021aun,Mikuni:2021nwn,Khosa:2022vxb,Letizia:2022xbe, Finke:2022lsu, Freitas:2022cno}. In particular, Ref.~\cite{Cranmer:2015bka} established that training a classifier and then ``calibrating'' it to learn its distribution under the relevant hypotheses guarantees a proper estimation of the Likelihood Ratio and thus of the optimal significance of an analysis. In the absence of systematic uncertainties, when the distribution of the classifier output is approximated by a binned Likelihood the resulting statistical model coincides with the previously detailed strategy employed by the ATLAS and CMS collaborations.

In this paper we propose a simplification of Ref.~\cite{Cranmer:2015bka} that can be used for any ensemble of events, combining the current ML-technique power to deal with high-dimensional data with the likelihood-based inference tests used in traditional analyses to discriminate between signal-plus-background and background-only hypotheses. Our method allows us to obtain the expected sensitivity when using these ML algorithms, both for discovery and for exclusion limits.

In order to assess the potential of our method, we will first consider a toy example in which we generate random data samples from multivariate Gaussian distributions. 
The motivation to do this is that we can compare the output of our method to the optimal classifier, which we can build since we know the actual generative functions.
As we will see, when facing low-dimensional problems, our method gives close-to-optimal results\footnote{We mostly focus on discovery sensitivities, although we provide all the relevant formulae for computing the exclusion limits.} and performs similarly to those obtained by following a standard binned Poisson log-likelihood approach.
On the other hand, while binning a multidimensional space becomes intractable, we will show that our method is still easy to apply when the dimensionality of the problem increases and, moreover, that it leads to results that are closer to the optimal classifier than those obtained by fixing a working point.
Finally, and as a more practical example, we will apply our method to a realistic problem of searching for heavy $W'$ bosons at the LHC, where we show that the statistical power of the analysis benefits greatly from the implementation of the method.

The paper is organized as follows: in Section~\ref{sec:method} we detail the method proposed to enhance statistical tests through ML classifiers, discussing in Section~\ref{sec:method-comparison} the differences between our method and previous ones proposed in the literature; Section~\ref{sec:apps} is dedicated to the application of this method to two examples, a toy model consisting of two Gaussians in varying dimensions and a realistic example extracted from the LHC Olympics datasets~\cite{Kasieczka:2021xcg}; finally, Section~\ref{sec:concl} is left to summarize our results and discuss future improvements.

\section{Method}\label{sec:method}

Our method combines the power of current ML techniques, see {\it e.g.} Ref.~\cite{Bishop:998831} for a very pedagogical introduction, to deal with high-dimensional data with the likelihood-based inference tests used in traditional analyses to discard different hypotheses~\cite{Cowan:2010js}. It is aimed as a different way to incorporate ML techniques to supervised searches for different Standard Model and BSM processes.

Suppose we have a set of $N$ independent measurements, each of which consisting of an arbitrarily high-dimensional set of observables $x$.\footnote{In the context of collider physics this could refer to kinematical variables such as $p_T$, $\eta$, invariant mass, etc.} We are interested in modelling the likelihood $\mathcal{L}$ of the data as a function of a background process $b$, a signal process $s$ and a signal strength parameter $\mu$, which defines the hypothesis we are testing for: a background-only hypothesis corresponds to $\mu = 0$ while the background-plus-signal hypothesis corresponds to $\mu = 1$. 
This likelihood function is nothing more than the probability of obtaining a given dataset conditioned on the aforementioned information and parameters:
\begin{equation}
    \mathcal{L}(\mu,s,b) = p(N,\{x_{i}, i=1,...,N\}|\mu,s,b)\,.
\end{equation}
A choice of likelihood function is a choice of a specific statistical model of the data. Following Ref.~\cite{cowan1998statistical}, we define the statistical model of $N$ independent measurements using the extended Likelihood
\begin{equation}
    \mathcal{L}(\mu,s,b) = \text{Poiss}\big(N|\mu S + B\big)\,\prod_{i=1}^{N}p(x_{i}|\mu,s,b)\label{eq:stat_model} \,,
\end{equation}
where $S$ ($B$) is the expected total signal (background) yield, Poiss stands for a Poisson probability mass function $\text{Poiss}(n|\lambda)=\frac{e^{-\lambda}\lambda^{n}}{n!}$ and $p(x|\mu,s,b)$ is the probability density for a single measurement $x$. 
Looking at this equation, there is an interplay between local and global information. The global Poisson term reflects the ensemble factor while $p(x)$ encodes the event-by-event information. The latter is the one that could be enhanced by ML analyses.

We can model the probability density as a mixture of signal and background densities
\begin{equation}
    p(x|\mu,s,b) = \frac{B}{\mu S + B}\,p_{b}(x)+\frac{\mu S}{\mu S + B}\,p_{s}(x)\,,\label{eq:prob_single_measurement}
\end{equation}
where $p_{s}(x)=p(x|s)$ and $p_{b}(x)=p(x|b)$ are, respectively, the signal and background probability densities for a single measurement $x$, and $\frac{\mu S}{\mu S + B}$ and $\frac{B}{\mu S + B}$ are the probabilities of an event being sampled from said probability densities.

Having defined a statistical model, we can follow Ref.~\cite{Cowan:2010js} and define the relevant test statistic $\tilde{t}_{\mu}$:
\begin{equation}
\tilde{t}_{\mu} =   \begin{cases}
    -2\text{ Ln }\dfrac{\mathcal{L}(\mu,s,b)}{\mathcal{L}(\hat{\mu},s,b)}       & \quad \text{if } \hat{\mu} \geq 0\,,\\[2ex]
    -2\text{ Ln }\dfrac{\mathcal{L}(\mu,s,b)}{\mathcal{L}(0,s,b)}  & \quad \text{if } \hat{\mu} < 0\,,
  \end{cases}
\end{equation}
where $\hat{\mu}$ is the parameter that maximizes the likelihood $\mathcal{L}(\mu,s,b)$. By differentiating Eq.~(\ref{eq:stat_model}) with respect to $\mu$ and finding its zeroes one can show that $\hat{\mu}$ is such that
\begin{equation}
    \sum_{i=1}^{N}\frac{p_{s}(x_{i})}{\hat{\mu}S\, p_{s}(x_{i}) +B\, p_{b}(x_{i})} = 1 \,.\label{eq:muhat_0}
\end{equation}
Notice that we recover the traditional counting experiment result $\hat{\mu}=\frac{N-B}{S}$ when the $x$ offers no discrimination power between $b$ and $s$, which implies $p_{s}=p_{b}$.

With the test statistic $\tilde{t}_{\mu}$, we can study the expected discovery potential and the expected upper limits of the analysis. The discovery potential corresponds to studying the background-only hypothesis $\mu=0$, where the test statistic $q_{0}\equiv\tilde{t}_{0}$ takes the form
\begin{equation}
q_{0} =   \begin{cases}
    -2\text{ Ln }\dfrac{\mathcal{L}(0,s,b)}{\mathcal{L}(\hat{\mu},s,b)}       & \quad \text{if } \hat{\mu} \geq 0\,,\\
    \phantom{-}0  & \quad \text{if } \hat{\mu} < 0\,,
  \end{cases}\label{eq:test-discovery}
\end{equation}
and plugging  Eq.~(\ref{eq:stat_model}) explicitly in
\begin{equation}
q_{0} =   \begin{cases}
    -2\hat{\mu}S + 2\sum_{i=1}^{N}\text{Ln }\left(1+\dfrac{\hat{\mu}S}{B}\dfrac{p_{s}(x_{i})}{p_{b}(x_{i})}\right)      & \quad \text{if } \hat{\mu} \geq 0\,,\\
    \phantom{-}0  & \quad \text{if } \hat{\mu} < 0\,.
  \end{cases}
  \label{eq:testdiscovery1}
\end{equation}

In general $p_{s,b}(x)$ are not known and are usually approximated by discrete binned distributions. For $D$ bins, one obtains in each bin $d$ the expected number of background events $B_{d}$, the expected number of signal events $S_{d}$ and the measured number of events $N_{d}$, so Eq.~(\ref{eq:stat_model}) turns to~\cite{Cranmer:2021urp}
\begin{equation}
    \mathcal{L}(\mu,s,b) = \prod_{d=1}^{D}\text{Poiss}\big(N_{d}|\mu S_{d} + B_{d}\big)\,.\label{binned-likelihood}
\end{equation}
This binned log-likelihood approximation is very effective but runs into trouble when the dimensionality of the data grows, as the finite statistics renders the density estimation unreliable. 
For this reason, we propose a different way of dealing with the high-dimensional dataset. 
We train a classifier to distinguish between the signal and background hypotheses with a balanced large dataset\footnote{Notice that this does not aim to reflect the measured set of $N$ events, as at this point, we are interested in estimating only the densities.}, obtaining a classification score $o(x)$ that maximizes the binary cross-entropy (BCE) and thus approaches
\begin{equation}
    o(x) = \frac{p_{s}(x)}{p_{s}(x)+p_{b}(x)}\,,
    \label{optimal-classifier}
\end{equation}
as the classifier approaches its optimal performance, see {\it e.g.}~the Machine Learning Chapter in Ref.~\cite{Workman:2022ynf}. This means that the classifier learns the per-instance likelihood ratio $\frac{p_{s}(x)}{p_{b}(x)}$, precisely the information needed in Eq.~(\ref{eq:testdiscovery1}). We can then reduce the dimensionality by dealing with $o(x)$ instead of $x$, using
\begin{equation}
    p_{s}(x) \rightarrow \tilde{p}_{s}(o(x))\,, \hspace{1cm} \text{and} \hspace{1cm} p_{b}(x) \rightarrow \tilde{p}_{b}(o(x))\,,
\end{equation}
where $\tilde{p}_{s,b}(o(x))$ are the distributions of $o(x)$ for signal and background, obtained by evaluating the classifier on a set of pure signal or background events, respectively. 
Notice that this allows us to approximate both signal and background distributions individually, although only the ratio will be relevant for estimating the expected sensitivities.
Since these distributions are one-dimensional, they can be easily binned and incorporated into Eq.~(\ref{eq:stat_model}). Therefore the test statistic of Eq.~(\ref{eq:testdiscovery1}) becomes
\begin{equation}
q_{0} =   \begin{cases}
    -2\hat{\mu}S + 2\sum_{i=1}^{N}\text{Ln }\left(1+\dfrac{\hat{\mu}S}{B}\dfrac{\tilde{p}_{s}(o(x_{i}))}{\tilde{p}_{b}(o(x_{i}))}\right)      & \quad \text{if } \hat{\mu} \geq 0\,,\\
    \phantom{-}0  & \quad \text{if } \hat{\mu} < 0\,,
  \end{cases}
  \label{eq:testdiscovery2}
\end{equation}
and the condition on $\hat{\mu}$ from Eq.~(\ref{eq:muhat_0})
\begin{equation}
    \sum_{i=1}^{N}\frac{\tilde{p}_{s}(o(x_{i}))}{\hat{\mu}S\, \tilde{p}_{s}(o(x_{i})) +B\, \tilde{p}_{b}(o(x_{i}))} = 1 \,.\label{eq:muhat}
\end{equation}

We shall name the resulting statistical model Machine-Learned Likelihood (ML Likelihood). In this sense, we are treating the algorithm as a dimensionality-reduction technique where we learn the appropriate one-dimensional manifolds that best discriminates between signal and background. This is different from the usual way of incorporating these algorithms to experimental analyses. We are neither assuming a working point and counting events selected by the algorithm in this working point nor interpolating the Likelihood as in Ref.~\cite{Coccaro:2019lgs}. Our method has a more concise goal which is to take advantage of the full information of the data in a supervised analysis by replacing the cut and count procedure for the likelihood-ratio information.

This method is a simplification of the one detailed in Ref.~\cite{Cranmer:2015bka} for likelihood-free inference, where we do not construct an unbinned likelihood ratio but use instead the ``calibrated'' estimated likelihoods obtained by applying density estimation techniques to the learned output function for each process. Although we are also using machine learning to reduce the dimensionality of the problem, we are taking an intermediate step where we only aim to approximate individual likelihoods and not to replace the likelihood-based test statistics with a learned, likelihood-free generalized log-likelihood ratio. This is evidenced by the fact that we exclude the signal strength from the training step, being instead a parameter to maximize in the manner detailed in Ref.~\cite{Cranmer:2015bka}. The simplification is possible because we are dealing with additive signal whose probability distribution does not depend on the signal strength. The parameterization of the Likelihood implemented here is in some sense analogous to the use of parameterized Likelihood Ratios for Effective Field Theory searches, see {\it e.g.}~Ref.~\cite{Chen:2020mev}. Our parameterization is even simpler but, as we show in Section~\ref{sec:apps}, still very useful to increase the statistical power of a given analysis.

The test statistic in Eq.~\eqref{eq:testdiscovery2} is estimated through a finite dataset of $N$ events and thus has a probability distribution conditioned on the true unknown signal strength $\mu'$. 
For a given hypothesis described by the $\mu'$ value, we can estimate numerically the $q_{0}$ distribution. With this distribution, one can estimate the median expected discovery significance $\text{med }[Z_{0}|\mu']$ by considering the median of the test statistic
\begin{equation}
    \text{med }[Z_{0}|\mu'] = \sqrt{\text{med }[q_{0}|\mu']}\,.
\end{equation}
In particular, in our results we will report the discovery significance of the signal-plus-background hypothesis $\text{med }[Z_{0}|1]$, where the significance encodes how likely is to the background-only hypothesis to explain data that follows the signal-plus-background hypothesis. A higher significance will thus imply that the background-only hypothesis can be excluded in favor of the signal-plus-background hypothesis with a larger confidence.

Notice that we do not introduce Asimov datasets here to provide an asymptotic estimation of the significance. This is because the introduction of $p(x)$ renders the definition of an Asimov dataset more complicated. 
We do instead a full numerical estimation where we generate a set of datasets generated under the signal-plus-background hypothesis and compute for each of them the test statistic $q_{0}$. 
Since our  method is relatively simple, a numerical estimation of the $q_{0}$ distribution is a feasible task. 
Indeed, this is an advantage of the one-dimensional representation of the data.

Nevertheless, since we will be interested in comparing our method to other standard techniques, we also introduce here the median discovery significance estimate for the binned likelihoods in Eq.~(\ref{binned-likelihood}) through the use of Asimov datasets, given by the well known formula~\cite{Cowan:2010js}:
\begin{equation}
    \text{med }[Z_{0}^{\text{binned}}|1] = \left[ 2 \, \sum_{d=1}^{D} \left( \left(S_{d} + B_{d} \right) \, \text{Ln}\left(1 + \frac{S_{d}}{B_{d}} \right) - S_{d} \right) \right]^{1/2}\,,
    \label{binned-Z}
\end{equation}
where again $B_{d}$ and $S_{d}$ are the expected number of background and signal events in bin $d$. 

For a realistic problem, the trained classifier is usually sub-optimal and the learned observable $o$ is an approximation of the log-likelihood ratio which may miss relevant information, thus reducing the power of the considered tests. 
Since we are not considering a specific working point but instead taking advantage of the full information retained in $o$, the degree of classification power of a classifier is captured by global metrics such as the Area-Under-Curve (AUC). 
The AUC is the integral of the Receiver Operating Characteristic (ROC) curve $\epsilon_{s/b}(\text{WP})$, where $\epsilon_{s/b}(\text{WP})$ are the fraction of correctly classified signal/background events as a function of the WP, with a higher AUC signaling a higher overall performance. 
The closer the AUC is to its largest possible value (which usually is below 1), the better $o$ captures the full distributions' information. Because of this, a higher AUC correlates with a larger significance, with an upper limit set by the optimal classifier. 
We emphasize that our method works regardless of whether the classifier is optimal or not, with optimality providing an upper limit of the test performance (which holds regardless of the method considered). 
The usefulness of this method is that it approximates the true likelihoods better than a binned log-likelihood analysis and thus provides a larger significance.

For completeness, we also provide the relevant steps to derive upper limits on $\mu$.
In this case, we need to consider the test statistic~\cite{Cowan:2010js}:
\begin{equation}
\tilde{q}_{\mu} =   \begin{cases}
    0       & \quad \text{if } \hat{\mu} > \mu\,,\\
    -2\text{ Ln }\frac{\mathcal{L}(\mu,s,b)}{\mathcal{L}(\hat{\mu},s,b)}       & \quad \text{if } 0 \leq \hat{\mu} \leq \mu\,,\\
    -2\text{ Ln }\frac{\mathcal{L}(\mu,s,b)}{\mathcal{L}(0,s,b)}  & \quad \text{if } \hat{\mu} < 0\,,
  \end{cases}
\end{equation}
and to look at its median expected significance when the true hypothesis is assumed to be the background-only one:
\begin{equation}
    \text{med }[Z_{\mu}|0] = \sqrt{\text{med }[\tilde{q}_{\mu}|0]}\,,
\end{equation}
where we estimate the $\tilde{q}_{\mu}$ distribution by generating a set of datasets with background-only events. 
Then, to set upper limits to a certain level, we select the lowest $\mu$ which achieves the required median expected significance.

Finally, we would like to mention that we have neglected the different systematic uncertainties that arise when performing any measurement. It is important to notice that one should also include in Eq.~(\ref{eq:stat_model}) a set of nuisance parameters $\theta$ to capture these systematic uncertainties, so $S$, $B$, $p_{s}$ and $p_{b}$ will be functions of these parameters. 
Extending our method with systematics could be relatively straightforward, with problems potentially arising when obtaining the ML Likelihood. 
The reason for this issue is that we need to compute ML Likelihood by training a ML algorithm, so dealing with these systematic errors requires some ingenuity. 
A possibility is to extend our training dataset from $x$ to $(x,\theta)$ with $\theta$ sampled from a prior distribution $p(\theta|x')$ from any additional measurements $x'$, in line with the treatment detailed in Ref.~\cite{Cranmer:2015bka,Ghosh:2021roe}. 
For the sake of simplicity, we will not include them in this analysis and leave them for future works.

\subsection{Similar approaches in the literature}
\label{sec:method-comparison}

There have already been several approaches to marry ML classifiers and statistical tests, see e.g.~\cite{Cranmer:2015bka, Elwood:2018qsr,DAgnolo:2018cun,Nachman:2019dol,DAgnolo:2019vbw,Chen:2020mev,Matchev:2020wwx,Cornell:2021gut,Aguilar-Saavedra:2021utu,dAgnolo:2021aun,Mikuni:2021nwn,Khosa:2022vxb,Letizia:2022xbe, Finke:2022lsu, Freitas:2022cno}. To our knowledge, the most similar methods to the current proposal can be found in Refs.~\cite{DAgnolo:2018cun,DAgnolo:2019vbw,Chen:2020mev,dAgnolo:2021aun,Khosa:2022vxb,Letizia:2022xbe}, with the latter two appearing during the completion of this work. Although we share several aspects, there are enough differences that warrant this proposal. 

In Refs.~\cite{DAgnolo:2018cun,DAgnolo:2019vbw,dAgnolo:2021aun,Letizia:2022xbe}, the authors propose a method to detect deviations from a reference dataset (namely, the SM). They parameterize the alternative hypothesis in terms of a learnable function, a Neural Network in Refs.~\cite{DAgnolo:2018cun,DAgnolo:2019vbw,dAgnolo:2021aun} and a non-parametric kernel in Ref.~\cite{Letizia:2022xbe}, which is trained to quantify discrepancies between the data and the reference model. The function then provides the log-likelihood ratio between the two hypotheses, which can then be used for hypothesis testing to discard the reference hypothesis. This is a very powerful tool for anomaly detection, but is not exactly what we are proposing in this work. We consider a supervised search where the two hypotheses are well defined and we are reducing the hypothesis test to a single parameter of interest, the signal strength $\mu$. Our method is thus simpler and easier to implement because we are not asking the algorithm to learn the dataset but only to learn the discriminator between two different processes in a high-dimensional space. The trade-off is a lack of flexibility and, at least in this form, an impossibility to perform a model-agnostic search.

On the other hand, Ref.~\cite{Khosa:2022vxb} states a similar goal to what we  present in this work: to obtain the significance of a supervised search that incorporates ML classifiers. They also extend the method to unsupervised searches, which is something our method is currently not designed to do. However, we note that the proposed statistical model and thus the questions that the statistical tests can answer are different. While our formulation in terms of statistical mixture models is an enhancement of traditional analyses, their likelihood proposal intends to differentiate between different types of ensembles. The authors obtain the output distribution for the background-only and for the signal-plus-background cases, while we obtain the background-only and signal-only distributions and introduce them as part of the mixture model. The question we aim to answer is thus different: while Ref.~\cite{Khosa:2022vxb} aims to discriminate between different types of ensembles, we intend to discriminate between different possible compositions of a single measured ensemble of data. The statistical test obtained reflects this difference. In our case, the test statistic is a simple extension of the usual methodology while in their case it is a different test. An additional advantage of our method is that we are reducing the problem to a single parameter of interest which we can then study.

The search for optimal sensitivity through parameterization in this work relates it to Ref.~\cite{Chen:2020mev}. There, the parameterization occurs at the Likelihood ratio level and is specific to the Effective Field Theory scenario, see also Refs.~\cite{Brehmer:2018eca,Brehmer:2018kdj} for a detailed explanation of the method. Parameterizing $\frac{p_{s}}{p_{b}}$ explicitly in terms of the parameters of interest, the Wilson coefficients, the learning task changes from learning the Likelihood Ratio to learning a set of functions which combined with the parameters of interest yield the Likelihood Ratio. The authors of Ref.~\cite{Chen:2020mev} show that this parameterization provides optimal discriminating power for a wide range of possible values of the parameters of interest without specific retraining. Our method is different in the sense that it tackles a different physics scenario, additive resonant physics as opposed to non-resonant EFT, and thus considers a different parameterization. The additive resonant physics scenario implies that we learn a mixture model where the parameter of interest is the signal strength $\mu$, which we do by estimating the individual likelihoods in the learned one-dimensional embedding space. This is in contrast to Ref.~\cite{Chen:2020mev} where the unbinned Likelihood ratio is learned by parameterizing it in terms of the known dependence on the parameters of interest at parton level.
\section{Applications}\label{sec:apps}

To explore the Machine-Learned Likelihood approach that connects ML classifiers with the standard statistical tests, in this section we compare the expected discovery significance $Z_{0}$ estimated following the method described in the previous section against the usual and naive counting experiment result. 
First we consider a toy model, where the data is generated from Gaussian variables. 
In this simple example, we can explore the potential and robustness of our ML Likelihood approach by comparing its performance against the optimal log-likelihood ratio statistical test obtained using the true underlying probability density functions (pdfs). 
Moreover, we also compare it with the calculation of $S/\sqrt{B}$ considering a subset of events obtained by applying different cuts, {\it i.e.} defining a working point, with the same classifier used in our estimation. 
Finally we study a more realistic situation, where the true generative functions are unknown, by considering a search for new BSM particles in a dijet final state at the LHC.

In both examples we train our per-event classifiers using {\tt XGBoost}~\cite{Chen:2016:XST:2939672.2939785}, an optimized gradient boosting library that provides a parallel tree boosting. Maximum depth was set to 5, the number of estimators up to 500, and {\tt binary:logistic} as objective to perform a logistic regression for binary classification. The evaluation metric for validation data is {\tt logloss}, and early stopping was established after 50 rounds to avoid overtraining. We have checked that modifying slightly the {\tt XGBoost} parameters does not change significantly our results. Furthermore, other ML algorithms suitable for the classification problem can also be used as long as they give good performance, for example deep neural networks. The use of Boosting algorithms for High Energy Physics is certainly not new, but its implementation for Likelihood estimation is not so common. This is probably due to the fact that its basis algorithm, Decision Trees, has been known to introduce non-smooth regions in the Likelihood Ratio estimator due to its very nature~\cite{Cranmer:2015bka}. However, the Boosting strategy circumvents that problem by the recursive application of Decision Trees. See Ref.~\cite{Chatterjee:2022oco} for a similar example of the power of Boosting Decision Trees for Likelihood Ratio estimation. In each scenario involving toy models 1M events per class where generated, while in the BSM analysis 100k signal and 100k background events were used. The training procedure was performed with half of the dataset available.

\subsection{Toy example with multivariate Gaussian distributions}

We begin with our first example involving events generated by multivariate Gaussian variables, $\mathcal{N}_{dim}(\boldsymbol m,\boldsymbol\Sigma)$, in several scenarios by increasing the dimensionality $dim=1,...,10$ of the problem. 
For each $dim$ we consider two multivariate Gaussian distributions with their covariance matrices fixed to the $dim\times dim$ identity matrix, $\boldsymbol\Sigma = \mathbb{I}_{dim\times dim}$, {\it i.e.} with no correlation between them, but with different means of $\boldsymbol m = +0.3$ $\mathds{1}_{dim} $ for the signal and $\boldsymbol m = -0.3$ $\mathds{1}_{dim}$ for the background, with $\mathds{1}_{dim}$ the size $dim$ vector of ones.

\subsubsection*{Dimension 2 case}

\begin{figure}[t!]
\begin{center}
 \begin{tabular}{c}
 \hspace*{-4mm}
 \epsfig{file=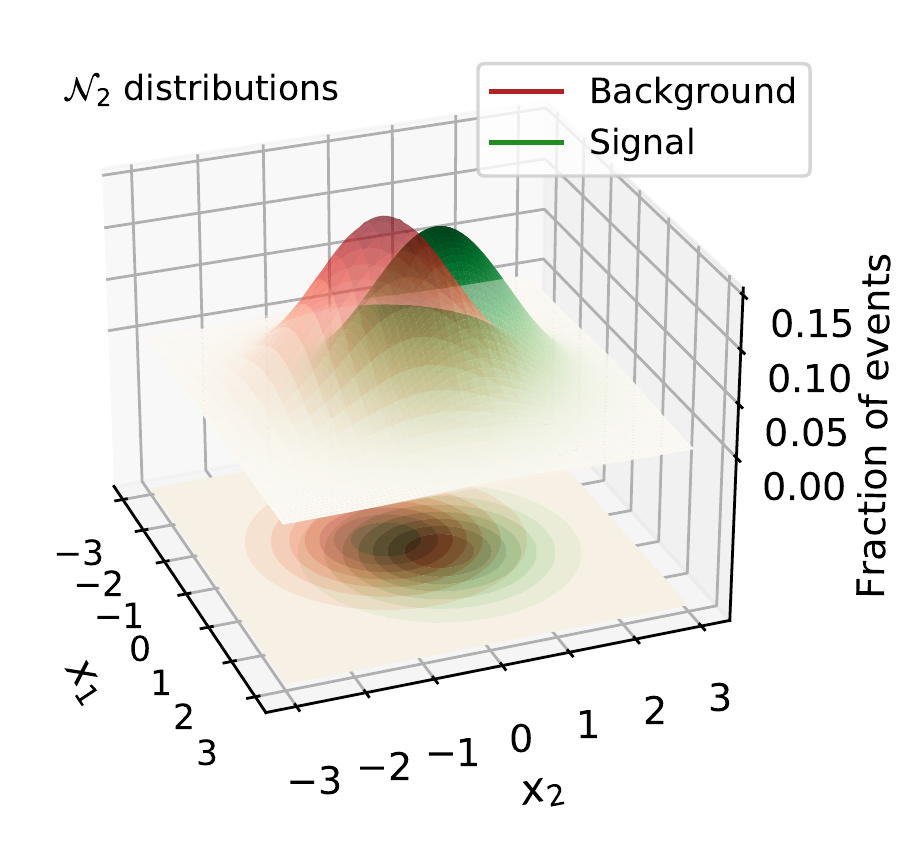,height=7cm} 
    \end{tabular}
    \caption{Multivariate Gaussian random variables, $dim=2$ example. $\mathcal{N}_{2}(+0.3$ $\mathds{1}_{2},\mathbb{I}_{2\times 2})(x)$ for signal, $\mathcal{N}_{2}(-0.3$ $\mathds{1}_{2},\mathbb{I}_{2\times 2})(x)$ for background.}
    \label{2gaussian-distributions}
\end{center}
\end{figure}

\begin{figure}[t!]
\begin{center}
 \begin{tabular}{cc}
 \hspace*{-5mm}
 \epsfig{file=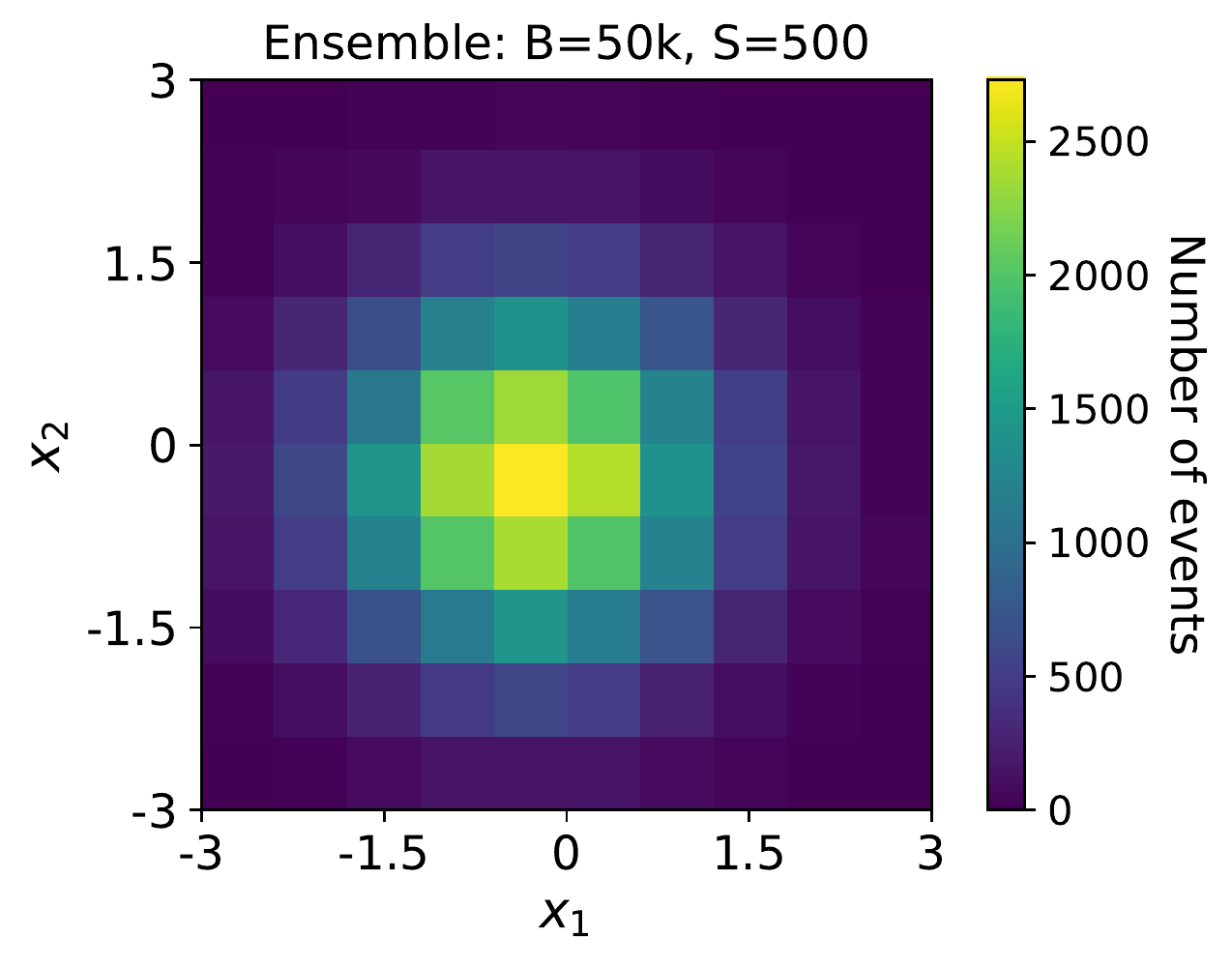,height=5.7cm} 
 \hspace*{-0mm} \epsfig{file=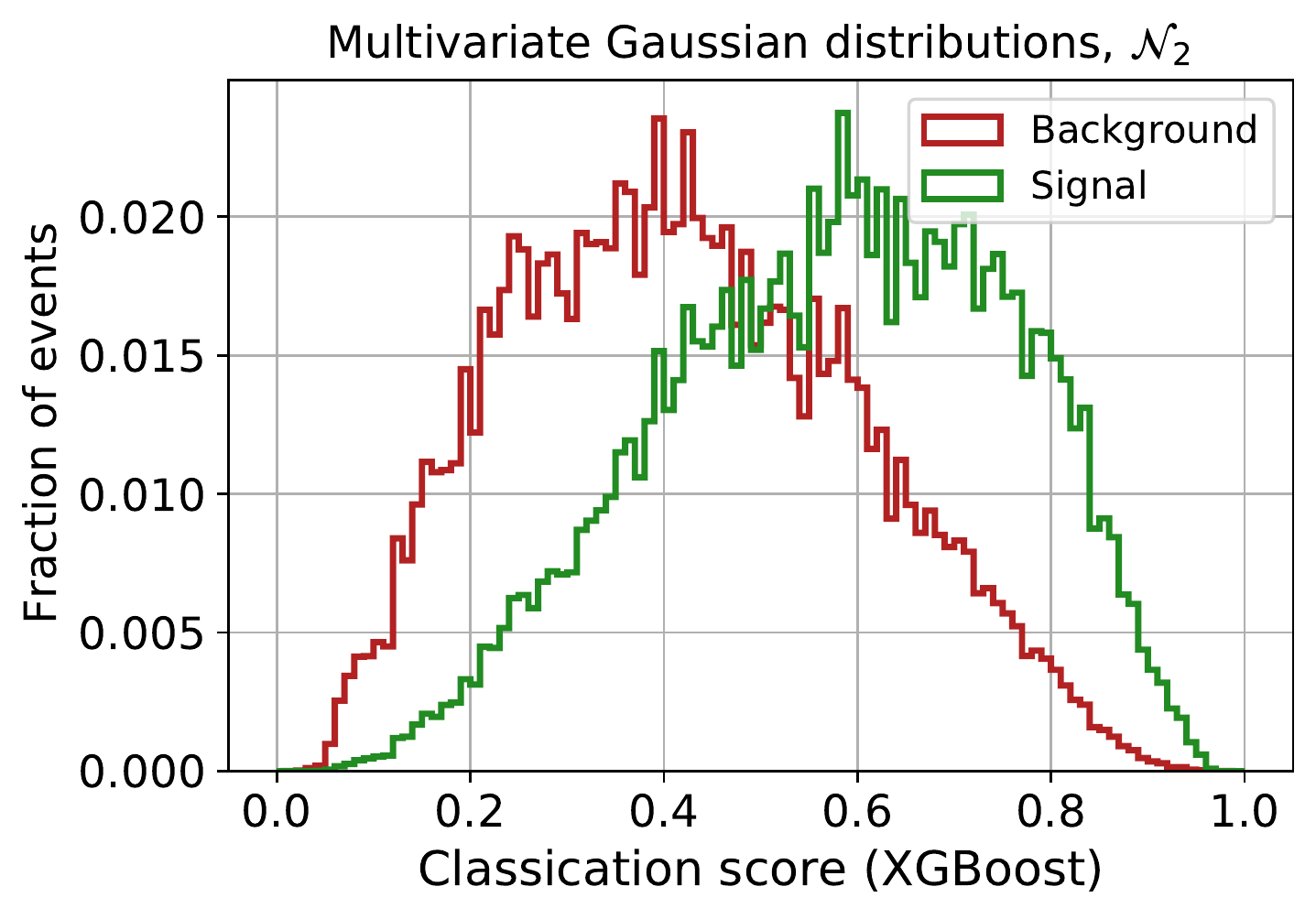,height=5.7cm} 
    \end{tabular}
    \caption{Left panel: histogram of an hypothetical experiment with $B=50$k and $S=500$ events divided into $10\times 10$ bins. Signal and background distributions can be seen in Fig.~\ref{2gaussian-distributions}. Right panel: classification score, $o(x)$, for a binary classifier using the {\tt XGBoost} algorithm.}
    \label{2gaussian-ensemble-output}
\end{center}
\end{figure}

To ease visualization,  we consider first $dim=2$, as shown in Fig.~\ref{2gaussian-distributions} with the signal  in green and background in red.
For concreteness, let us consider a fixed expected number of background events, $\langle B \rangle=50$k, and a free number of signals events, $\langle S \rangle$, that we vary to evaluate the performance of the ML Likelihood method. 
On the left panel of Fig.~\ref{2gaussian-ensemble-output} we show an example of how an ensemble with $S=500$ events would look like.
We stress again that this is just a toy model in abstract space $(x_1,x_2)$. 
In a real life experiment, such as a collider analysis, this could correspond for example to the transverse momentum and pseudorapidity of a jet, $x_1=p_T$ and $x_2=\eta$.
Moreover, the expected signal-to-background ratio would be set by the relative cross-sections and the total amount of events by the effective luminosity.

In order to apply the ML Likelihood method introduced in the previous section, we need to estimate the likelihood ratio $p_{s}(x)/p_{b}(x)$, which we obtain by training a supervised per-event classifier, {\tt XGBoost}. 
At this stage we are only interested in obtaining a classifier to distinguish between signal and background, therefore to train and test the algorithm we employ all the events in our dataset, {\it i.e.} a large and balanced sample. As usual, we label signal events with a 1 and background events with a 0. A histogram of the resulting classification score, $o(x)$, can be seen on the right panel of Fig.~\ref{2gaussian-ensemble-output} for two new independent datasets of pure signal (green) and pure background (red) events.
Then, we use these classification output distributions to estimate the per-instance likelihood ratio

\begin{equation}
    \frac{p_{s}(x)}{p_{b}(x)} \rightarrow \frac{\tilde{p}_{s}(o(x))}{\tilde{p}_{b}(o(x))}\,.
\end{equation}

This is a key step of the method, where we approximate the signal and background {\it dim}-dimensional pdfs, $p_{s,b}(x)$, by binning the 1-dimensional ML classification score $o(x)$, and thus taking advantage of its dimensionality-reduction power.

Now we can focus on the test statistic of Eq.~(\ref{eq:testdiscovery2}). Following the procedure of Section~\ref{sec:method}, first we need to compute $\hat{\mu}$, the signal strength value that maximizes the likelihood. To do this we construct 10k ensembles mixing background and signal events such that the number of events per class are taken from Poisson distributions with means $\langle B \rangle=50$k and a value of $\langle S \rangle$. For each ensemble, solving Eq.~(\ref{eq:muhat}) we obtain numerically a value of $\hat{\mu}$ that we finally use to calculate the test statistic $q_0$.

\begin{figure}[t!]
\begin{center}
 \begin{tabular}{c}
 \hspace*{-4mm}
 \epsfig{file=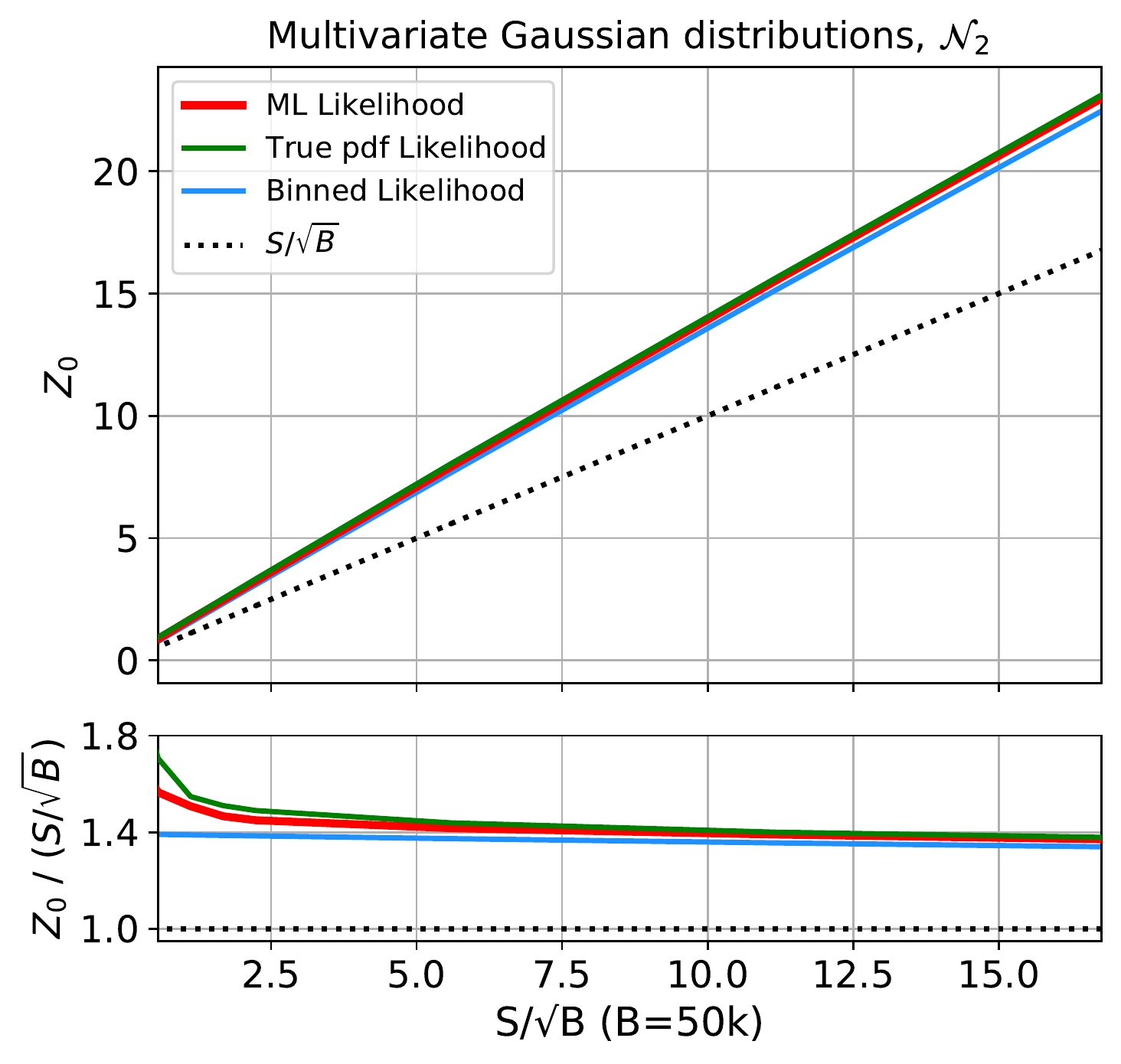,height=8cm} 
    \end{tabular}
    \caption{Discovery significance calculated with various methods for the example in Fig.~\ref{2gaussian-ensemble-output} for fixed background, $\langle B \rangle=50$k, and different signal strengths $\langle S \rangle$. 
Red lines show the results implementing the ML Likelihood method with {\tt XGBoost} used to estimate the probability density for single events and green lines when using the true multivariate distributions. 
The black dotted line represents the result of the usual counting method, $S/\sqrt{B}$, in the entire range of interest (only one bin), and the light blue curve the result of a binned counting experiment.}
    \label{2gaussian-Z}
\end{center}
\end{figure}

The median expected discovery significance for the ML Likelihood method, estimated as the median of the test statistic, is shown in Fig.~\ref{2gaussian-Z} as a red curve. 
Notice that $Z_{0}$ is depicted as a function of $S/\sqrt{B}$, therefore the identity relation (black dotted curve) represents the naive significance approximation of Eq.~(\ref{binned-Z}) considering a single bin and $S \ll B$.
 We can see that the curve is above this naive estimate even for low significances, $Z_{0}<1$. 

On the other hand, since we are dealing with a low dimensional problem, we can also employ a binned Poisson log-likelihood approximation, Eq.~(\ref{binned-likelihood}), and its median discovery significance estimated by introducing the Asimov dataset, Eq.~(\ref{binned-Z}). 
Then, we calculate numerically $B_{d}$ and $S_{d}$ constructing ensembles of $10 \times 10$ bins  with $\langle B \rangle=50$k and different values of $\langle S \rangle$ from a 1M events database of per class. 
The resulting median significance is also shown in Fig.~\ref{2gaussian-Z} as a light blue curve. 

Finally, we also present in Fig.~\ref{2gaussian-Z} as a green curve the significance using the true probability density functions, {\it i.e.} $p_{s,b}(x)=\mathcal{N}_{2}(\pm \boldsymbol m,\boldsymbol\Sigma)(x)$ for our example. Since we do not use a classifier to approximate $p_{s,b}(x)$, we can consider it as an optimal scenario and, thus, the green curve represents an upper limit for $Z_{0}$. 
For this simple example, we see that the results obtained using the ML Likelihood approach are very close to the optimal scenario and slightly outperforms the binned Poisson method.

Before moving to a higher dimensional scenario, let us explore the new method in more detail.
First of all, it is important to highlight that the classification score is one-dimensional by construction regardless of the dimensionality (the number of components) of our data and therefore can be easily binned.
This is not the case for the binned Poisson log-likelihood approximation, where the number of bins needed to estimate the density increases with the number of components and can eventually be problematic. 
In this $dim=2$ example each feature range on the left panel of Fig.~\ref{2gaussian-ensemble-output} is divided in 10 bins, therefore we end up with $10^{dim=2}=100$ bins. 
However, as the complexity, {\it i.e.} the dimension, of the problem grows, the problem of efficiently binning every feature becomes exponentially more challenging, and in practice intractable for a finite number of events.
On the other hand, with the ML Likelihood method we do not need to bin the original features, but the one-dimensional classification score, which was divided into 100 bins to approximate $\tilde{p}_{s,b}(o(x))$ as can be seen on the right panel of Fig.~\ref{2gaussian-ensemble-output}. The introduction of binning can reduce the obtained significance for the ML Likelihood as the Likelihood ratio $\tilde{p}_{s}(o(x))/\tilde{p}_{b}(o(x))$ is approximated to an averaged version, just as it does when going from the True Likelihood to the Binned Likelihood. However, we emphasize again that because we are binning in one dimension, optimal choices of binning can be easily explored.

\begin{figure}[t!]
\begin{center}
 \begin{tabular}{cc}
 \hspace*{-8mm}
 \epsfig{file=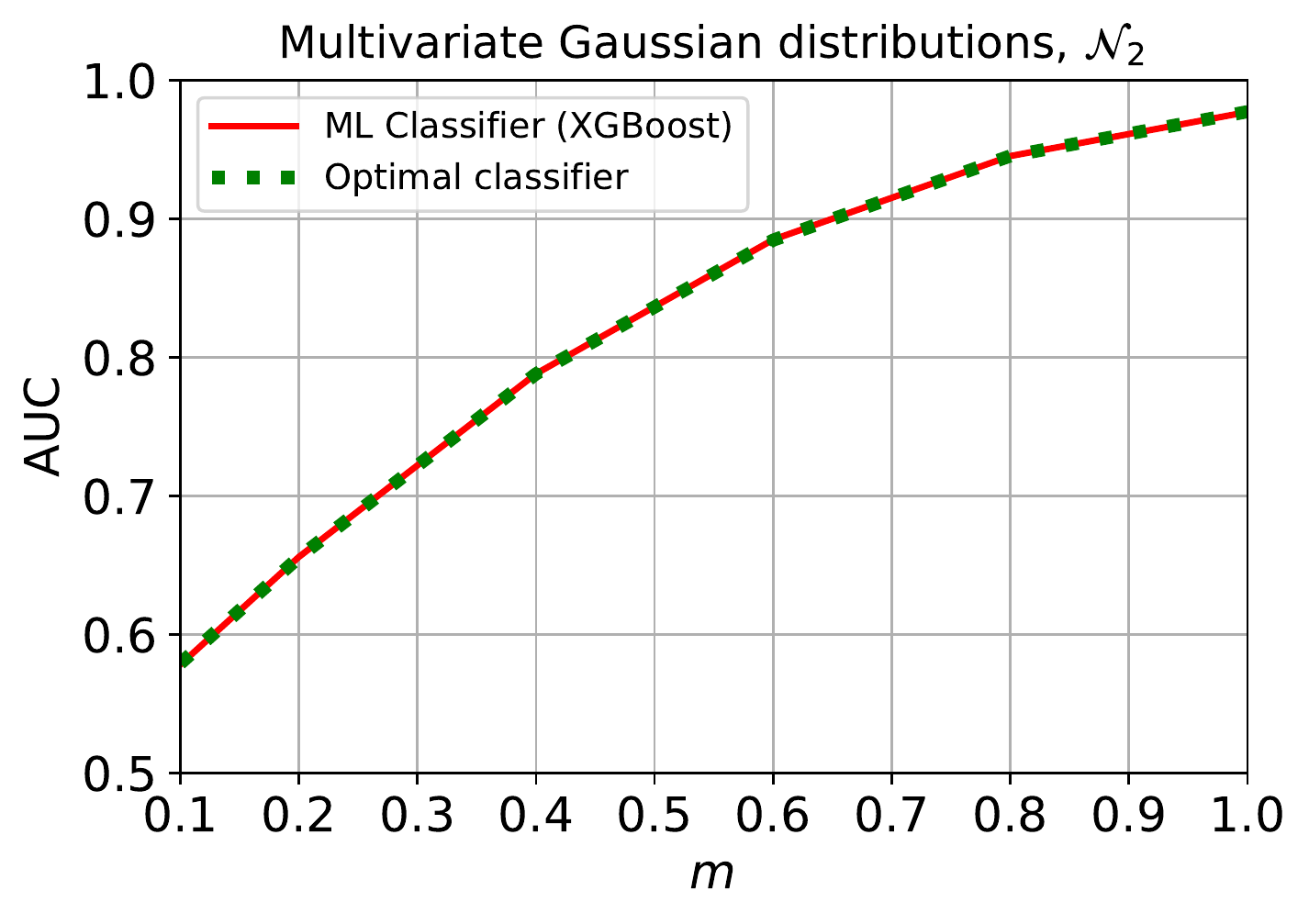,height=5.7cm} 
 \hspace*{-2mm} \epsfig{file=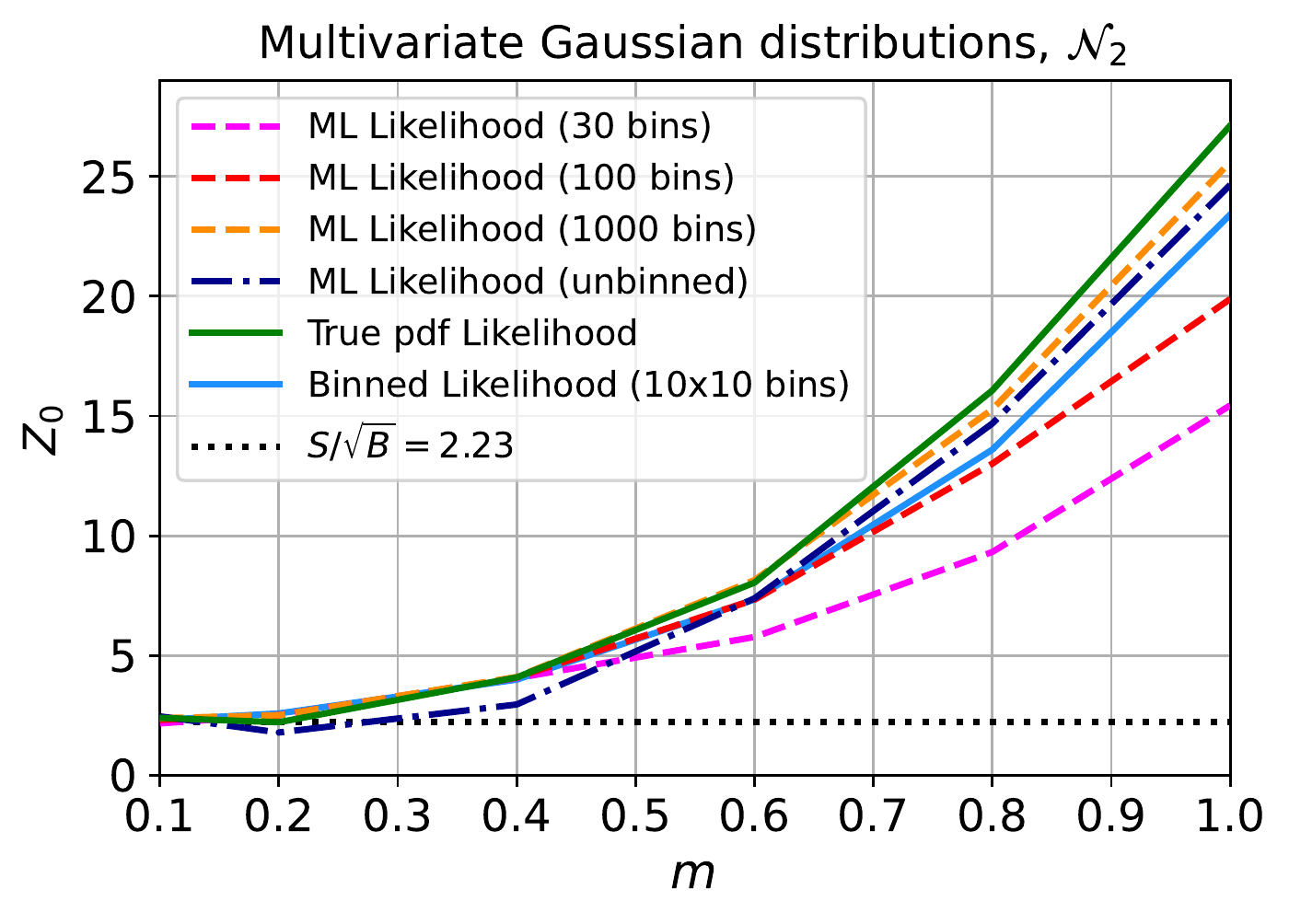,height=5.7cm} 
    \end{tabular}
    \caption{Left panel: AUC obtained with XGBoost (red curve) and the optimal classifier calculated with Eq.~(\ref{optimal-classifier}) (green dotted curve) for the multivariate Gaussian example of Fig.~\ref{2gaussian-ensemble-output}, but with increasing signal-background separation $m$. Right panel: discovery significance calculated with various methods for the same example. Here we fixed  $\langle B \rangle=50$k and $\langle S \rangle=500$. Color coding is the same as in Fig.~\ref{2gaussian-Z}, but we also include the results for the ML Likelihood method with different binning choices of the classification score and the significance obtained using the estimated unbinned Likelihood Ratio.}
    \label{2-gaussian-Z-mean}
\end{center}
\end{figure}

Second, we want to explore how the new method behaves when changing the performance of the classifier itself.
In order to do that, we still consider the same two $dim=2$ multivariate Gaussian distributions, but we vary their means, $m$.
Notice that increasing values of $m$ imply larger separation between signal and background, and therefore the classifier performs better. 
This is seen on the left panel of Fig.~\ref{2-gaussian-Z-mean}, where we display the classification power of the ML algorithm, measured by its AUC.
As expected, from the left and right panels of Fig.~\ref{2-gaussian-Z-mean} we can see that larger values of AUC imply higher significances. 

We also observe that the difference between the ML Likelihood significance and the True Likelihood significance increases with the AUC, even when the classifier is approximately optimal as seen in the left panel of Fig.~\ref{2-gaussian-Z-mean}. This is an effect of the suboptimal choice of binning. There is thus an additional approximation involved beyond the dimensionality reduction mentioned in Section~\ref{sec:method} that can reduce the significance of the ML Likelihood method compared to the True Likelihood significance. Recall that, if the classifier is sub-optimal, the ML Likelihood significance will be inevitably lower than the true significance as the dimensionality reduction causes information loss. Additionally, the ML Likelihood significance can be further reduced if the $o(x)$ binning is suboptimal in the sense of capturing the $p_{s}(o(x))/p_{b}(o(x))$ behavior appropriately. This is amenable by exploring optimal choices of one dimensional binning. In Fig.~\ref{2-gaussian-Z-mean} we include additional binning choices and observe a clear dependence on the ML Likelihood significance. From this we can be certain that the classifier is indeed optimal and thus any loss in significance can be attributed to the choice of binning. We also include the Significance obtained by using the classifier output to estimate directly the unbinned Likelihood Ratio $p_{s}(x)/p_{b}(x)$ through Eq.~\eqref{optimal-classifier} that is needed to estimate the test statistic detailed in Eq.~\eqref{eq:testdiscovery1}. We observe how the unbinned Likelihood Ratio performance lies between the 100 bins performance and the 1000 bins performance. If the unbinned Ratio was perfectly estimated, its performance should be identical to the true Likelihood performance. However, finite statistics of the training sample lead to imperfect estimation and, specially, to numerical uncertainties. The latter is one of the main motivations behind the introduction of calibration in Ref.~\cite{Cranmer:2015bka}. We thus show how our method, although binning dependent, yields a comparable performance to the unbinned method but with increased stability and robustness against numerical effects on the Likelihood Ratio estimation.

On the remainder of this work we consider the intermediate choice of 100 bins as it is a good compromise between approximating the optimal results and increasing computing costs. In realistic applications, the optimal AUC is not known a priori and thus one should explore binning choices to achieve maximum significance for a given AUC. A possible avenue is suggested by previous experimental analyses~\cite{ATLAS:2017bqn,ATLAS:2018tgm,ATLAS:2019fxb,CMS:2019uxx,CMS:2020poo,ATLAS:2020efs,ATLAS:2020pvn,CMS:2020uim,CMS:2021kom}, where the binning is decided in terms of a number of bins and a maximum statistical uncertainty per bin. 
An alternative is to implement other density estimation techniques such as Kernel Density Estimation or Normalizing Flows.

\begin{figure}[t!]
\begin{center}
 \begin{tabular}{c}
 \hspace*{-2mm} \epsfig{file=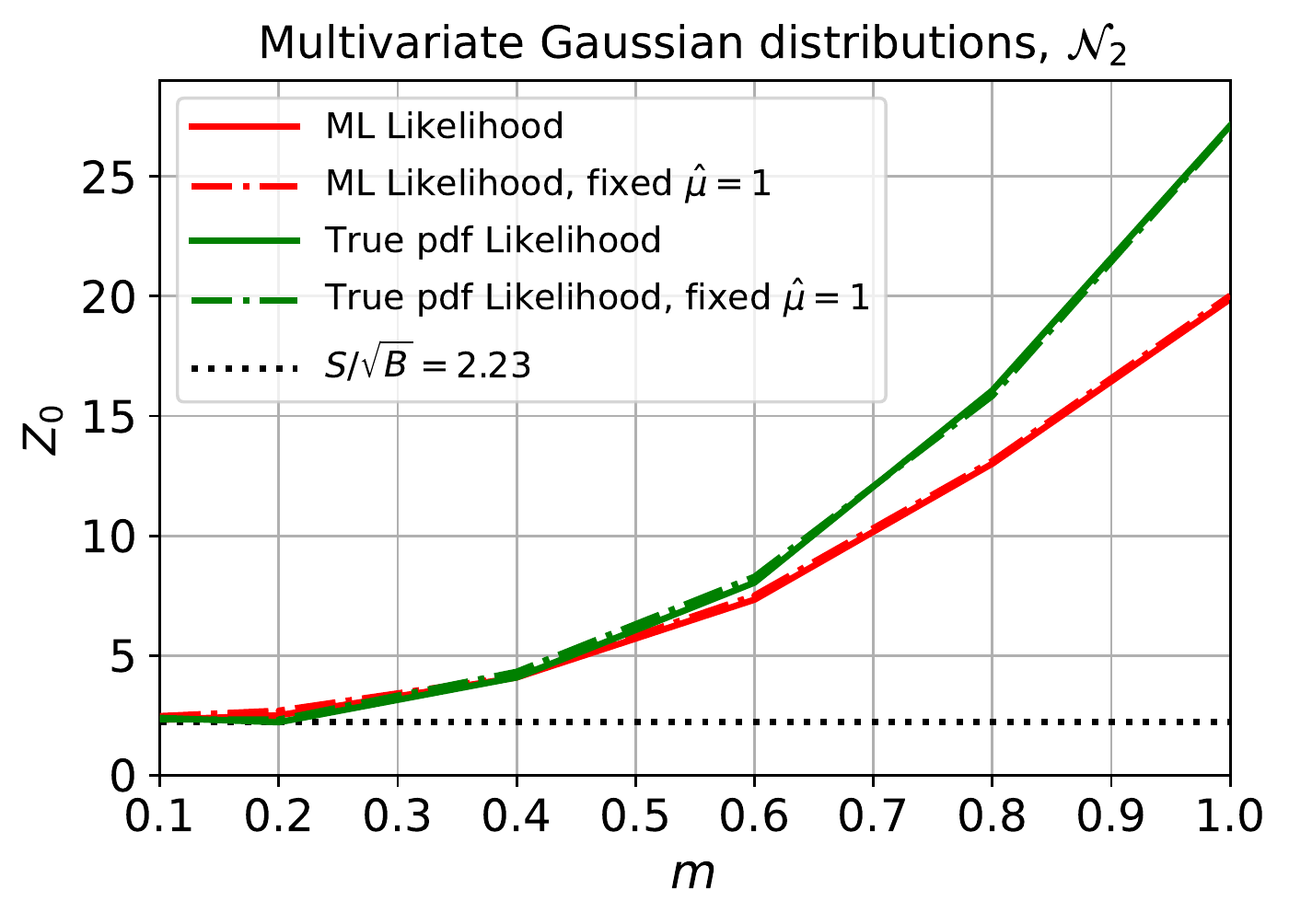,height=5.7cm} 
    \end{tabular}
    \caption{Discovery significance calculated with various methods for the same multivariate Gaussian example used in Fig.~\ref{2-gaussian-Z-mean}, increasing signal-background separation $m$, and fixed  $\langle B \rangle=50$k and $\langle S \rangle=500$. We compare the results when computing $\hat{\mu}$ numerically from data (solid) with those obtained after the approximation $\hat{\mu}=1$ (dot-dashed). Both ML Likelihood results use a classification score divided into 100 bins.}
    \label{2-gaussian-Z-mean-muhat}
\end{center}
\end{figure}

Finally, a practical comment about the implementation is in order.
Notice that we treat each value of $\langle S \rangle$ as a different hypothesis, thus we construct independent ensembles and compute $\hat{\mu}$ for each scenario. 
This can be a somewhat tedious task and could be the source of numerical errors in the significance estimation, so one possible simplification is to repeat the significance estimation procedure fixing $\mu=1$. The change in strategy between learning the parameter of interest and keeping it fixed amounts to change from a composite hypothesis test to a simple hypothesis test where the two hypothesis are fixed.
In Fig.~\ref{2-gaussian-Z-mean-muhat} we compare the results when $\hat{\mu}$ is calculated numerically from data satisfying Eq.~\eqref{eq:muhat}, and when we fix $\mu=1$, using both the true probability densities and the ML Likelihood. 
We see that the relative differences are small. This has a two-fold importance: it shows that the numerical procedure for $\hat{\mu}$ is correct, and also that the simple and composite hypotheses yield consistent significances, {\it i.e.}~that the simple hypothesis with no learned parameters of interest and the composite hypothesis where we estimate $\hat{\mu}$ have the same discriminating power. 
The latter is extremely useful to assess optimality, as the Neyman-Pearson Lemma that ensures optimal power for the Likelihood Ratio test is valid for simple hypotheses.
Notice moreover that this is true when using either the true Likelihoods or our ML-estimated Likelihoods, showing that our method yields an optimal test statistic that can be obtained using the one-dimensional learned output.
Indeed, we have checked that the simple and composite hypotheses are consistent for all the examples considered in this work both for the ML Likelihood and the true Likelihoods when available, although for brevity we only report the cases with estimated $\hat{\mu}$.

\subsubsection*{High-dimensional cases}

\begin{figure}[t!]
\begin{center}
 \begin{tabular}{cc}
 \hspace*{-8mm}
 \epsfig{file=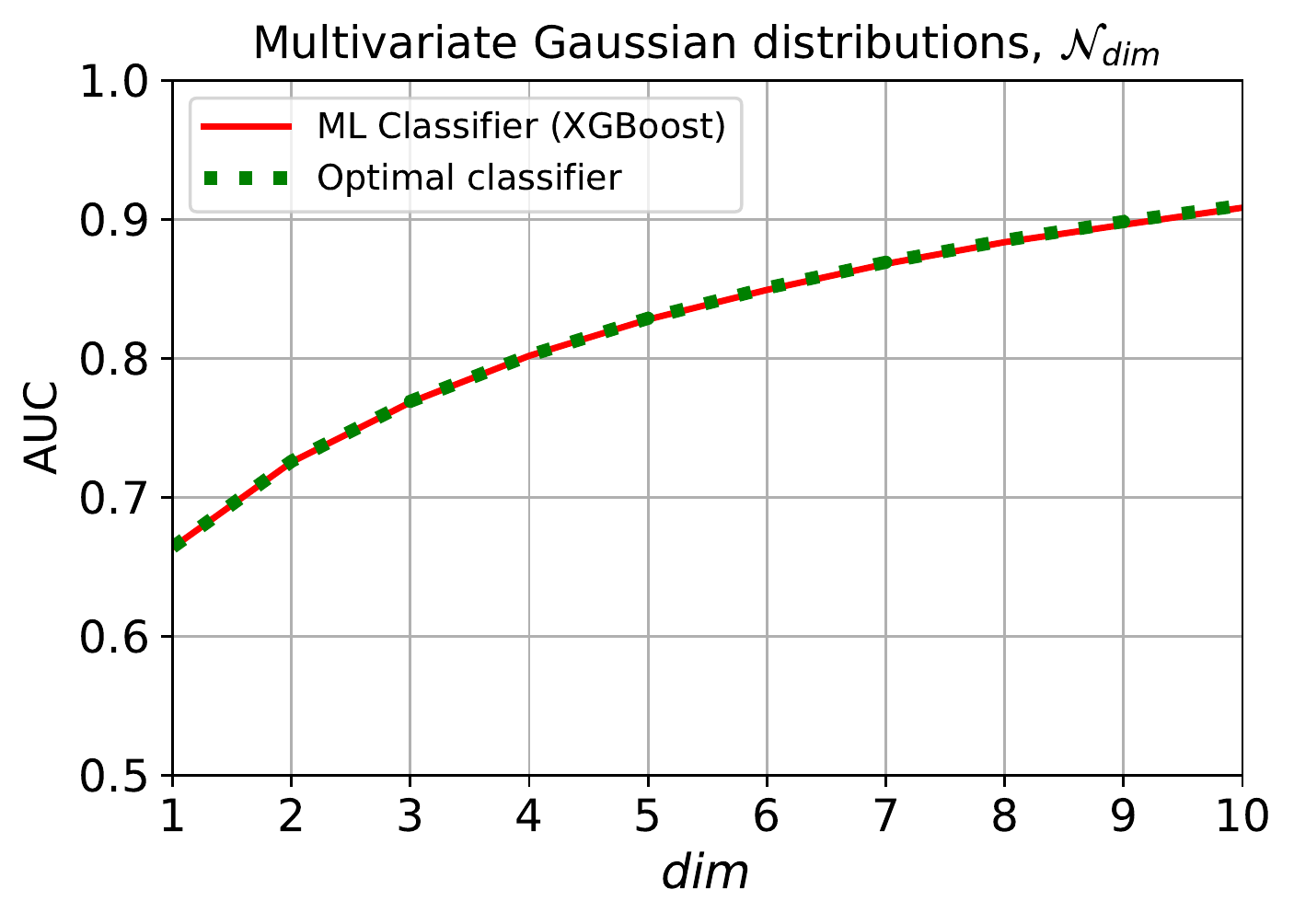,height=5.7cm} 
 \hspace*{-2mm} \epsfig{file=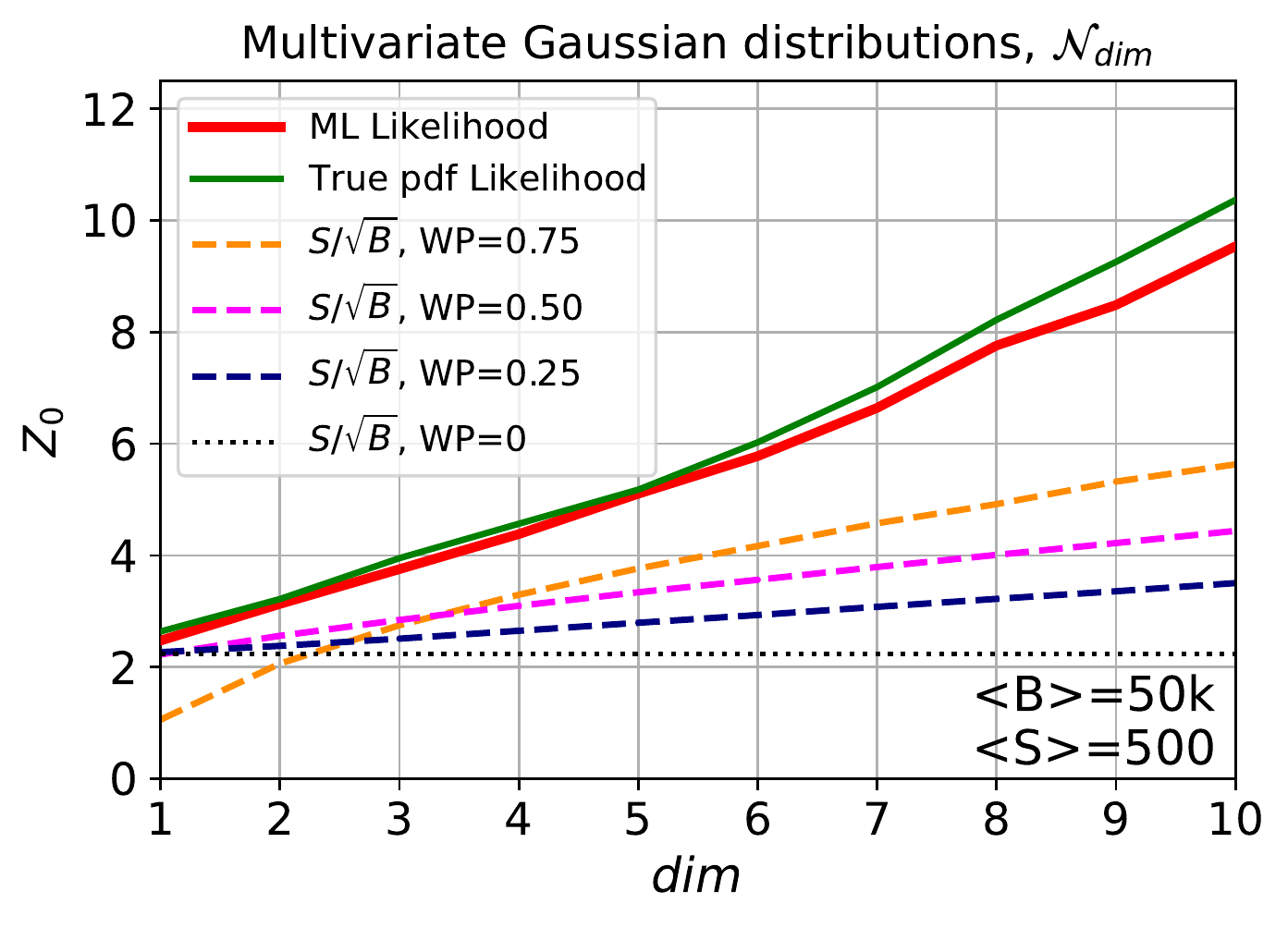,height=5.7cm} 
    \end{tabular}
    \caption{Left panel: AUC for the binary classifier using {\tt XGBoost} (red curve) and the optimal classifier calculated with Eq.~(\ref{optimal-classifier}) (green dotted curve) as a function of the data dimension, $dim$, for multivariate Gaussian variables. Right panel: significance calculated with various methods as a function of $dim$. For every case, the background and signal strengths were fixed, $\langle B \rangle=50$k and $\langle S \rangle=500$. Solid lines show the results implementing the method described in this work with: {\tt XGBoost} used to estimate the probability density for single events (red), the true multivariate distributions (green). The dashed curves represent the result of the usual counting method (only one bin, $S/\sqrt{B}$), but for a subsample of the original data found with {\tt XGBoost} assuming several working points, WP$=0.75, 0.5, 0.25$ to obtain signal enriched regions. The black dashed line also represents the result of the usual counting procedure, but considering the entire dataset (equivalent to WP$=0$).}
    \label{dimgaussians-AUC-Z}
\end{center}
\end{figure}

We repeat the procedure considering higher dimensional data, $\mathcal{N}_{dim}(\boldsymbol m,\boldsymbol\Sigma)$, with $dim=1, ..., 10$, $\boldsymbol\Sigma = \mathbb{I}_{dim \times dim}$, and $\boldsymbol m = +0.3$ $\mathds{1}_{dim} $ for the signal and $\boldsymbol m= -0.3$ $\mathds{1}_{dim}$ for the background. Notice that we treat each value of $dim$ as an independent hypothesis. The results are shown in Fig.~\ref{dimgaussians-AUC-Z}. 
On the left panel we present the AUC of the classifier and, as expected, it increases with the data dimensionality, since we introduce more information with each extra component making it increasingly easier to distinguish signal from background. We can also see that the classifier found is approximately optimal.
On the right panel we show the discovery significances obtained with several procedures for a fixed value of $\langle B \rangle=50$k and $\langle S \rangle=500$. 

As before, the green line is computed using the true pdf distributions, and therefore provides an upper limit for the performance of obtaining $Z_{0}$. 
The results of the ML Likelihood method, which can be easily computed also for higher dimensions, are shown in red. 
By comparing the two lines, we see that both have the same tendency, although the difference between the two increases for higher $dim$.
This can again be attributed to a suboptimal choice of $o(x)$ binning. The increase of dimensions produces the same behavior as the increase in separation for a fixed dimension (shown in Fig.~\ref{2-gaussian-Z-mean}): the two Gaussians are more distinguishable and the optimal AUC increases. Since the classifier is able to capture the increase in AUC, the $o(x)$ distribution gets more concentrated on the boundaries and thus the binning is not able to capture the likelihood ratio granularity as efficiently as for lower AUCs. Although it lies beyond the scope of this work, a ML Likelihood implementation in a real analysis where we do not know the true pdf can explore different binning choices. Even if suboptimal, ML Likelihood still provides a good estimate of the significance for relatively high dimensional problems.

Nevertheless, the main difference of increasing the dimensionality of the problem is the challenge it implies for computing the binned Poisson likelihood, as we did before for $dim=2$. 
For example, if every component range is divided in 10 we get $10^{dim}$ bins, which rapidly becomes intractable for a finite amount of statistics (we recall that each of our ensemble has $\sim 50$k events).
A commonly followed procedure to face this kind of situations is to used a ML algorithm as a classifier and to define a lower cut or working point in its output $o(x)$ to define signal enriched regions and to calculate $Z_{0}=S/\sqrt{B}$ on the resulting subset. 
We do this using the already trained {\tt XGBoost} classifier and defining several WP, including the particular case of WP=0 equivalent to applying no cut.
The results of this method are shown as dashed lines in the right panel of Fig.~\ref{dimgaussians-AUC-Z}.
For this example, we see that increasing the value of the WP does help improving the significances, and for values of WP=0.75 they approach to our results from the ML likelihood.
Notice that this is an interesting result, since the ML Likelihood method was able to perform better, {\it i.e.} closer to the optimal green line, without the need of defining and optimizing a WP. 
This is due to the fact that it makes use of the complete output of the classifier, thus including as much information as possible. In addition to including all WPs, this method has the advantage that it does not need to lose events by defining a more exclusive signal region. For relatively low total number of events, this is an important enhancement.

\subsection[Realistic application: a $W^{\prime}$ study at the LHC]{Realistic application: a $\boldsymbol{W^{\prime}}$ study at the LHC}

In this subsection we will focus on a collider physics example taken from the LHC Olympics~\cite{Kasieczka:2021xcg} challenge. The database is comprised by dijets events from two different sources: SM quantum chromodynamics (QCD) processes (background), and the production of a BSM new resonance $W'$ with mass $m_{W'}=3.5$ TeV.
This new particle decays to two new particles $X$ and $Y$ with masses $m_{X}=500$ GeV and $m_{Y}=100$ GeV, which in turn both decay promptly to a pair of quarks producing two large-radius jets with two-prong substructure (signal). The selected events have a reconstructed dijet mass within $[3.3,3.7]$ TeV.

\begin{figure}[t!]
\begin{center}
 \begin{tabular}{c}
 \hspace*{-4mm}
 \epsfig{file=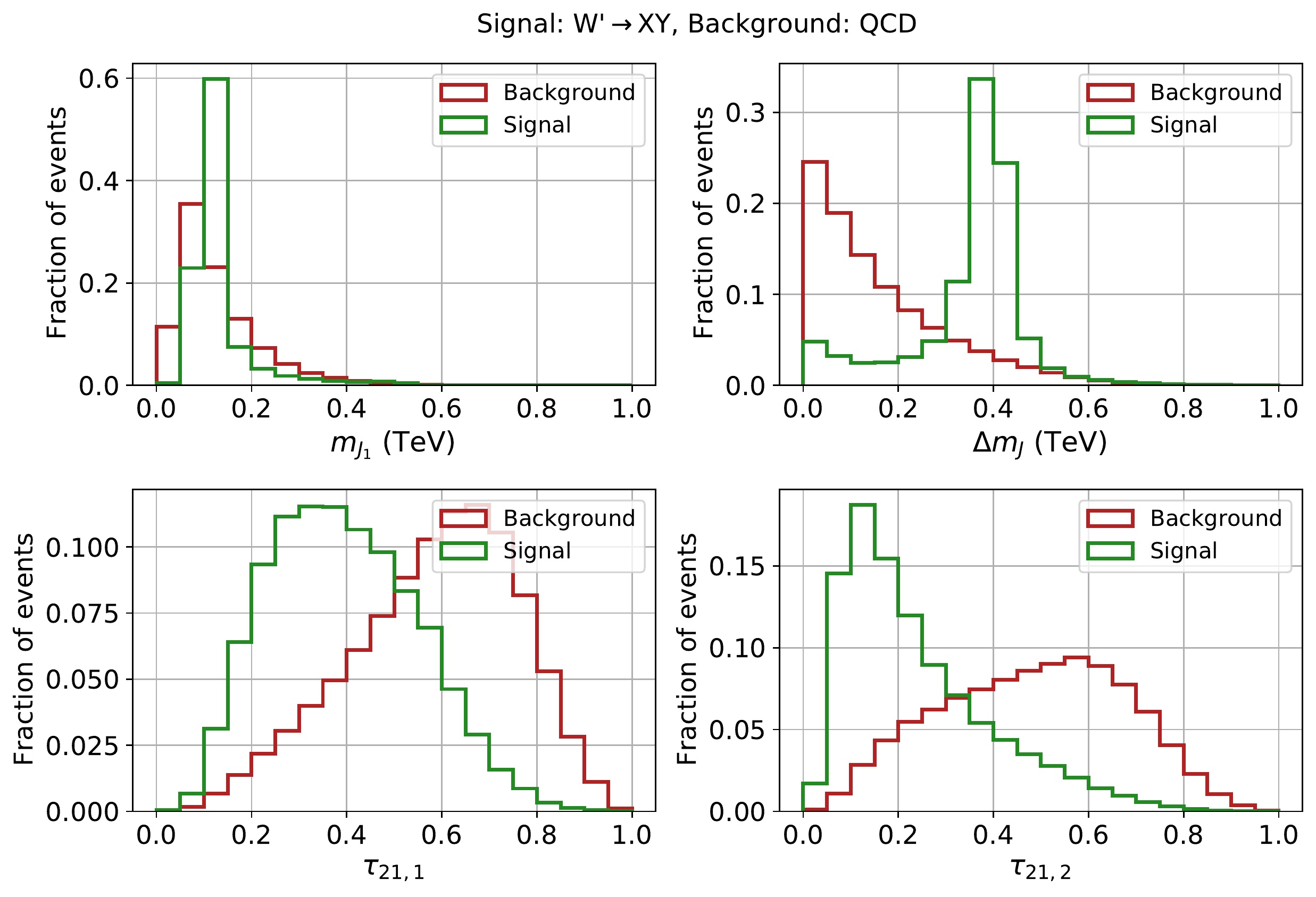,height=10.5cm} 
    \end{tabular}
    \caption{Distribution of the four features that characterize the dijet events. Signal corresponds to a new resonance $W' \rightarrow XY$ both decaying to two large-radius jets with two-prong substructure, and background to QCD dijet processes. The signal distributions on the top panels are centered at $m_{Y}=100$~GeV (left) and $m_X-m_{Y}=400$~GeV (right), while on the bottom panels the lower values $\tau_{21}$ signal distributions indicate a two-prong substructure.}
    \label{Wprime-variables-plot}
\end{center}
\end{figure}

Four features are considered to characterize the process and used to test the ML Likelihood method (thus we are dealing with a $dim=4$ problem): the invariant mass of the lighter jet ($m_{j1}$), the mass difference of the leading two jets ($\Delta m_j$), and the N-subjettiness ratios of the leading two jets ($\tau_{21,1}$ and $\tau_{21,2}$) \cite{Thaler:2010tr,Thaler:2011gf}. The latter parameters quantify if a jet is described by one or two subjets, indicating a two-prong substructure for smaller values. In Fig.~\ref{Wprime-variables-plot} we show the distributions of these parameters for the signal and background database. 

\begin{figure}[t!]
\begin{center}
 \begin{tabular}{cc}
 \hspace*{-8mm}
 \epsfig{file=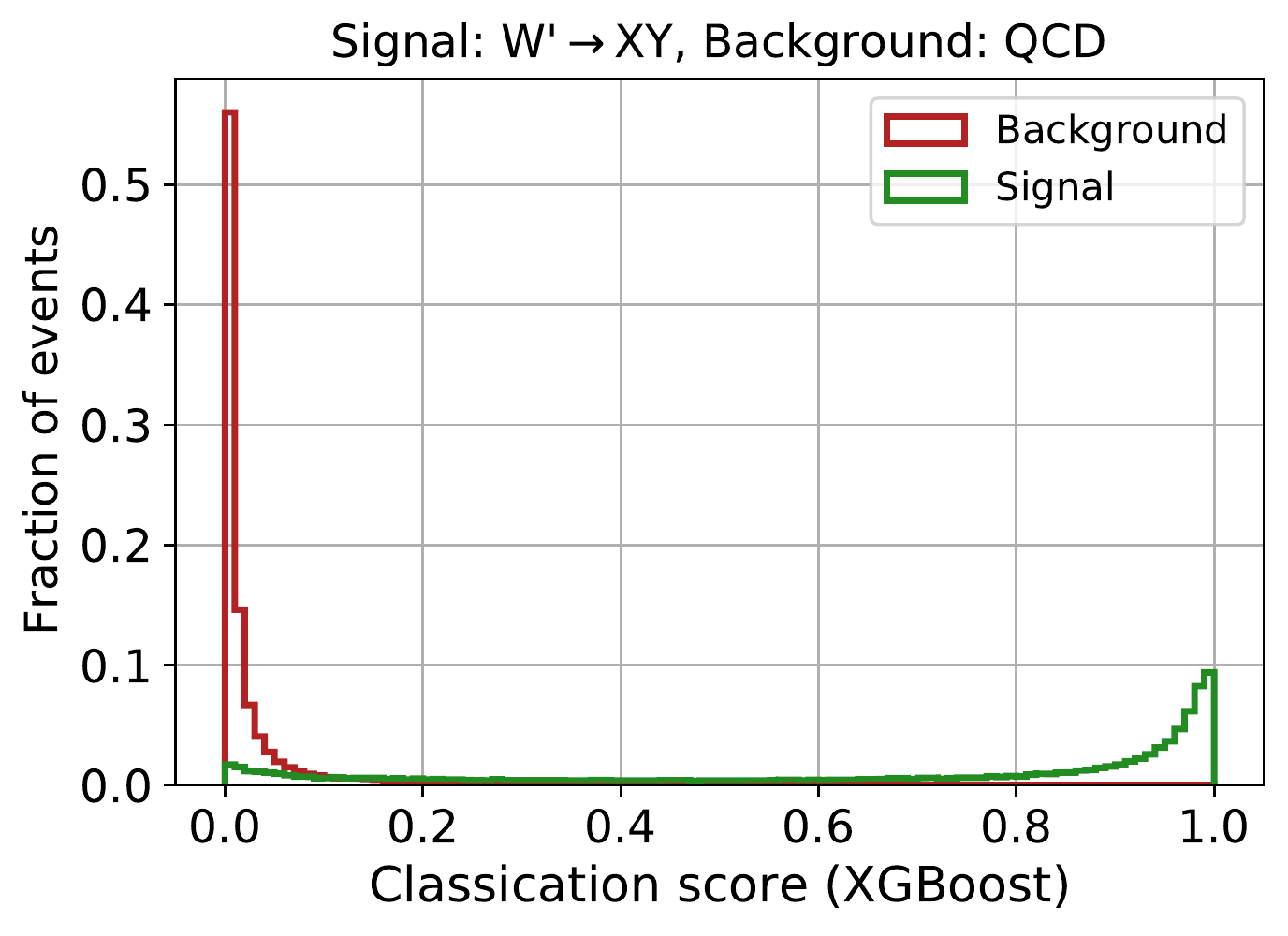,height=5.7cm} 
 \hspace*{-2mm} \epsfig{file=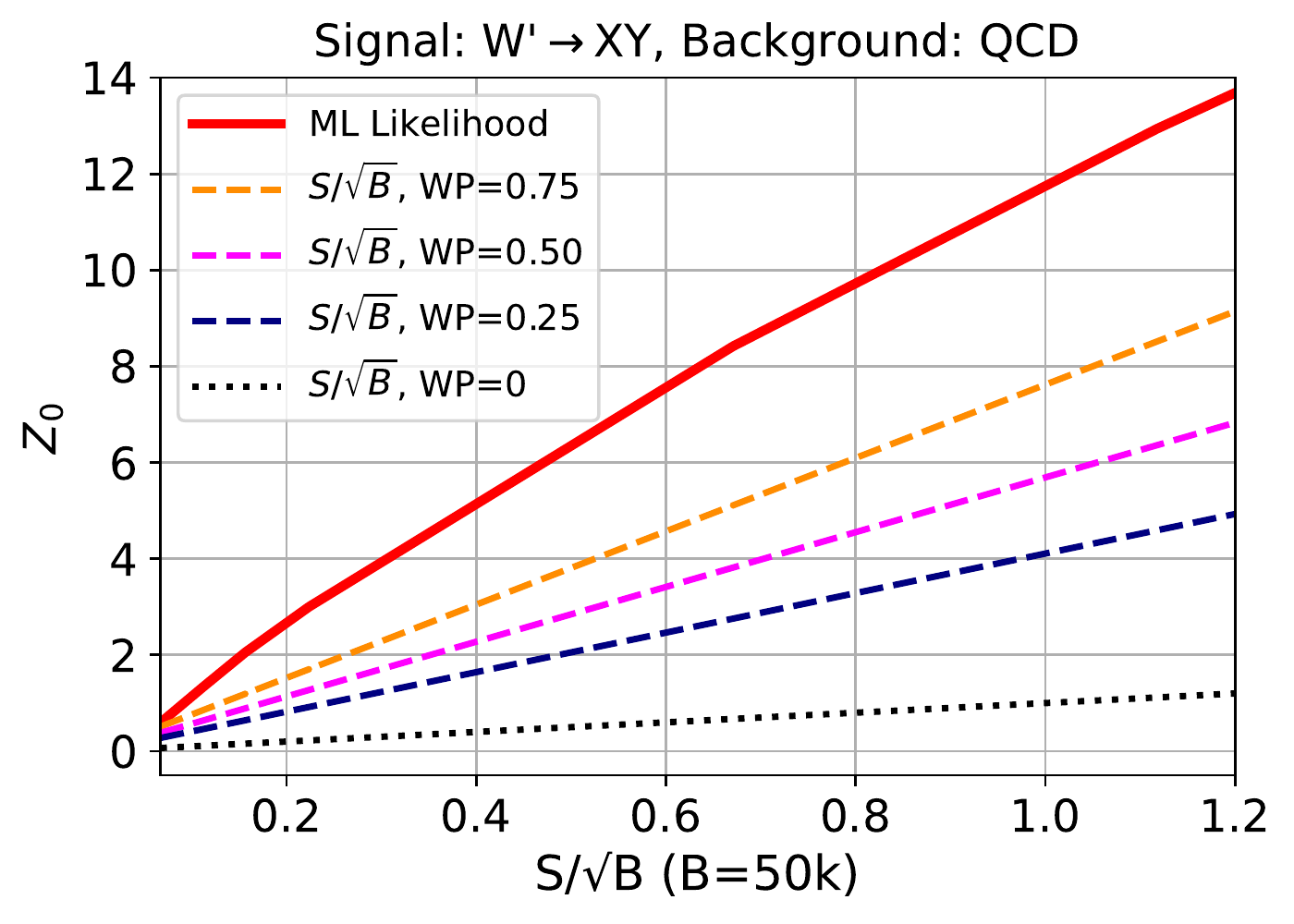,height=5.7cm} 
    \end{tabular}
    \caption{Left panel: classification score for the BSM search of $W'$ using a {\tt XGBoost} binary classifier. Independent pure signal and pure background test samples are evaluated to estimate $p_{s,b}(x)$. Right panel: significance calculated with several methods as a function of the signal-to-background ratio for $\langle B \rangle=50$k and $\langle S \rangle$ within $[10,300]$. Same color coding as in Fig.~\ref{dimgaussians-AUC-Z}. A 5$\sigma$ discovery significance could be found for $S \gtrsim 85$ with our method, and for $S \gtrsim 151, 200, 272, 1118$ with a usual counting method on enriched signal regions obtained with the same classifier but WP$=0.75, 0.5, 0.25, 0$, respectively.}
    \label{Wprime-output-signif}
\end{center}
\end{figure}

As in the multivariate Gaussian example, we train {\tt XGBoost} with the same hyper parameters to obtain a per-event binary classifier. We obtain AUC$=0.96$, meaning that the algorithm can distinguish between signal and background very efficiently, as can be seen on the left panel of Fig.~\ref{Wprime-output-signif} where the classification score, $o(x)$, is shown divided into 100 bins. Then, we estimate $p_{s,b}(x)$ by the discrete binned distribution of the ML output: $\tilde{p}_{s,b}(o(x))$. We would like to highlight again that we are binning a one-dimensional distribution, while in the usual binned Poisson approximation a four-dimensional space (the number of features that describes the process) would have to be binned.

To illustrate the method we fixed the expected number of background events $\langle B \rangle=50$k and vary the expected signal events $\langle S \rangle$ within the range $[10,300]$, since the signal-to-background ratio value would be determined by a particular model. We obtain numerically $\hat{\mu}$ satisfying Eq.~(\ref{eq:muhat}) by constructing 10k ensembles mixing $B$ and $S$ events taken from Poisson distributions with means $\langle B \rangle=50$k and different fixed $\langle S \rangle$. We obtain the ML Likelihood test statistic of Eq.~(\ref{eq:testdiscovery2}) for each value of $\langle S \rangle$, and finally estimate the median expected discovery significance as the median of the test statistic. The results are shown on the right panel of Fig.~\ref{Wprime-output-signif} as a red curve.

We also use the already trained {\tt XGBoost} classifier to calculate the significance by a traditional method. We consider the following working points to define a subsample within an enriched signal region, WP$=0.75,0.5,0.25$. We count only the events that satisfy $o(x)>$WP and compute $S/\sqrt{B}$, shown as dashed lines on the right panel of Fig.~\ref{Wprime-output-signif}. The special case WP$=0$ shown as a black dotted line represent the naive counting significance $S/\sqrt{B}$ in the entire range, i.e. using all the events. Comparing the results it is clear that the ML Likelihood method exceeds the usual ones. For example a 5$\sigma$ discovery significance would be obtained for $S \gtrsim 85$, while the number of expected signal events needs to be $ \gtrsim 151, 200, 272, 1118$ for WP$=0.75, 0.5, 0.25, 0$, respectively.

\section{Discussion and outlook}\label{sec:concl}

In this paper we have developed a simple method, called Machine-Learned Likelihood (ML Likelihood), which can be used for any ensemble of events, that combines the current ML-technique power to deal with high-dimensional data, with the likelihood-based inference tests used in standard analyses to discriminate between different hypotheses in a minimal way, which makes it amenable for exploratory analyses without high computational costs. 
It allows to obtain the expected experimental sensitivity when using ML algorithms, both for discovery and exclusion limits, evidencing the utility of these sort of algorithms even for resonance searches where the parameter of interest is a signal strength. 

Unlike other methods, the one proposed here makes use of all the output of the classifier, taking advantage of the entire ROC curve, and therefore its performance is better described by global quantities such as the AUC. 
Nevertheless, it is a rather simple method, as it is based on a single classifier that is used to estimate the individual probability densities and then to evaluate the significance in the statistical test given the values of $S$ and $B$ for the considered ensemble, in the manner of the calibrated classifiers proposed in Ref.~\cite{Cranmer:2015bka}. 
Notice that this is actually a key ingredient of the method, since the output of the classifier is one-dimensional by construction, and therefore the method has the advantage of always being easy to bin or even fit accurately, irrespectively of the actual dimensionality of the problem at hand.

In order to test the potential of the proposed method, we have applied it to two cases: a toy model with data generated from Gaussian variables, and a $W^\prime$ search in dijet final states at the LHC. 
The former was particularly relevant to understand the performance of the method, as we were able to compare the results obtained using the ML-estimated pdf against the true generative ones. 
We found that the new method leads to results that are close to the optimal case and, as expected, they remain so also for high-dimensional problems.
Moreover, while its performance is similar to traditional binning analysis for low-dimensional problems, we saw that the ML Likelihood method is particularly effective for more complex problems, where traditional binning is no longer possible and standard ML analysis are used by defining a particular working point.
This improvement has been found also in the more realistic analyses of $W^\prime$ searches at LHC, where the true generative functions are unknown. 
Again, we obtain higher significances with the ML Likelihood method. 

To summarize, we proposed a simple method to estimate statistical significances when using ML classification algorithms. 
It has the main advantages of remaining simple and reliable also for high dimensional problems, and of making use of the full knowledge of the ML algorithm, without the need of relying on a given working point for the analysis.
Yet, we have seen that it leads to excellent results, approaching the optimal ones computed with the true generative functions, and improving those obtained by traditional analysis techniques.

Finally, it should be mentioned that the main lacks of the ML Likelihood method are that it is not valid for anomaly detection nor can be applied to unsupervised analyses. 
Besides, we have not incorporated systematic uncertainties in the calculation of significance presented here. 
All these issues are relevant for any real analysis and will thus be addressed in future publications.

\vspace{2.5mm}
\paragraph{Acknowledgments.}
The authors thank Martin de los Rios and Rosa María Sandá Seoane for useful discussions. This work is partially supported by the ``Atracci\'on de Talento'' program (Modalidad 1) of the Comunidad de Madrid (Spain) under the grant number 2019-T1/TIC-14019 (EA), by the Spanish Research Agency (Agencia Estatal de Investigaci\'on) through the Grant IFT Centro de Excelencia Severo Ochoa No CEX2020-001007-S, funded by MCIN/AEI/10.13039/501100011033 (EA, XM), and by CONICET and ANPCyT under projects PICT 2017-2751, and PICT 2018-03682 (EA, AM, AP, AS).
XM acknowledges partial financial support also by Grant PID2019-108892RB-I00 funded by MCIN/AEI/10.13039/ 501100011033 and by the European Union’s Horizon 2020 research and innovation programme under the Marie Sk{\l}odowska-Curie grant agreement No 860881-HIDDeN. 
VML acknowledges the financial support by the Deutsche Forschungsgemeinschaft (DFG, German Research Foundation) under Germany's Excellence Strategy – EXC 2121 ``Quantum Universe'' – 39083330 and grant María Zambrano UP2021-044 funded by Ministerio de Universidades and ``European Union-NextGenerationEU/PRTR''.
MS acknowledges the financial support from the Slovenian Research Agency (grant No.~J1-3013 and research core funding No.~P1-0035).


\bibliographystyle{JHEP} 
\bibliography{biblio}

\end{document}